\documentclass[aps,twocolumn,showpacs,superscriptaddress,prl,10pt]{revtex4-1}
\usepackage{graphicx}
\usepackage{dcolumn}
\usepackage{bm}
\usepackage{color}
\usepackage{amsmath}
\usepackage{amsfonts}
\usepackage{subfigure}
\usepackage[makeroom]{cancel}

\newcommand{\COMMENTED}[1]{}

\begin{document}

\author{Ettore Vitali}
\affiliation{Department of Physics, The College of William and Mary, Williamsburg, Virginia 23187}
\affiliation{Department of Physics, California State University Fresno, Fresno, California 93740
}

\author{Hao Shi}
%\affiliation{Department of Physics, The College of William and Mary, Williamsburg, Virginia 23187}
\affiliation{Center for Computational Quantum Physics, Flatiron Institute, 162 5th Avenue, New York, New York 10010}

\author{Adam Chiciak}
\affiliation{Department of Physics, The College of William and Mary, Williamsburg, Virginia 23187}

\author{Shiwei Zhang}
\affiliation{Department of Physics, The College of William and Mary, Williamsburg, Virginia 23187}
\affiliation{Center for Computational Quantum Physics, Flatiron Institute, 162 5th Avenue, New York, New York 10010}

\title{Metal-insulator transition in the ground-state of the three-band Hubbard model at half-filling} 

\begin{abstract}
The three-band Hubbard model is a fundamental %minimal 
model for understanding properties of the Copper-Oxygen planes in cuprate superconductors. 
We use cutting-edge auxiliary-field quantum Monte Carlo (AFQMC) methods to investigate  
ground state properties of the model in the parent compound. %the charge transfer energy.
Large supercells 
combined with twist averaged boundary conditions are studied to reliably reach the thermodynamic 
limit. Benchmark quality results are obtained on the magnetic correlations and charge gap.
A key parameter of this model is the charge-transfer energy $\Delta$ between the Oxygen $p$  and 
the Copper $d$ orbitals, which appears to vary significantly across 
different families of cuprates and whose \emph{ab initio} determination is subtle. 
We show that the system 
undergoes a quantum phase transition from an antiferromagnetic insulator to 
a paramagnetic metal as $\Delta$ is lowered to $ 3\,$eV.
\end{abstract}

%\pacs{02.70.Ss, 71.27.+a, 71.10.Fd, 03.75.Ss}
% \keywords{}
\maketitle

It is widely believed that the physical mechanism underlying
high-temperature superconductivity in the cuprate materials
lies in the quasi-two-dimensional physics of the 
CuO$_2$ 
planes.
A significant amount of 
the theoretical %approaches
studies
of such planes 
 (see, e.g., Refs.~\cite{RevModPhys.87.457,RevModPhys.78.17} for some recent reviews)
have relied on the celebrated Hubbard Hamiltonian \cite{Hubbard_1,Hubbard_2},
which is a minimal low-energy effective model
that assumes the explicit contribution
of the Oxygen degrees of freedom can be %somehow 
neglected.
Although impressively accurate results \cite{hubbard_benchmark} have been obtained on the one-band Hubbard 
model and very interesting magnetic and charge orders have emerged \cite{Zheng1155,Mingpu-sc-PRB} which are relevant to some important experimental results, it is still unclear whether the model can support long-range superconducting correlations
in the ground state. Indeed the most recent and accurate numerical results seem to indicate that the answer is likely negative. 
While this answer in the one-band Hubbard model (or perhaps the closely related 
$t$-$J$ model which could contain different physics 
\cite{Sorella_tJ,Plakida_tJ,Sachdev_tJ,Bejas_tJ}) is clearly important and of fundamental value, 
it is timely, based on current results,
 to revisit what the effect of additional realism is and what might be a more 
accurate minimal model of the CuO$_2$  plane.

With the advent of modern
computing platforms
and progress in the development of numerical methods, it is now possible to reach beyond the
one-band %Hubbard 
model in favor of the more realistic, although still
minimal, three-band Hubbard model, also called the Emery model \cite{PhysRevLett.58.2794},
and obtain computational results of high accuracy and sufficiently close to the thermodynamic limit.
In this work, we 
perform an extensive study of the ground state of this model for the parent compounds,  employing %relying on 
the cutting-edge %determinantal Constrained Path Quantum Monte Carlo (CPQMC) 
constrained-path  auxiliary-field quantum Monte Carlo (CP-AFQMC)
method \cite{cpprl1995,AFQMC-lecture-notes-2013},
together with recently developed self-consistency loops \cite{PhysRevB.94.235119} to systematically improve 
the approximation needed because of the fermion sign problem.
The method maintains polynomial computational complexity, and we study large supercells under
twisted boundary conditions to determine properties at the thermodynamic limit.  

\begin{figure}[ptb]
\begin{center}
\includegraphics[width=9.0cm, angle=0]{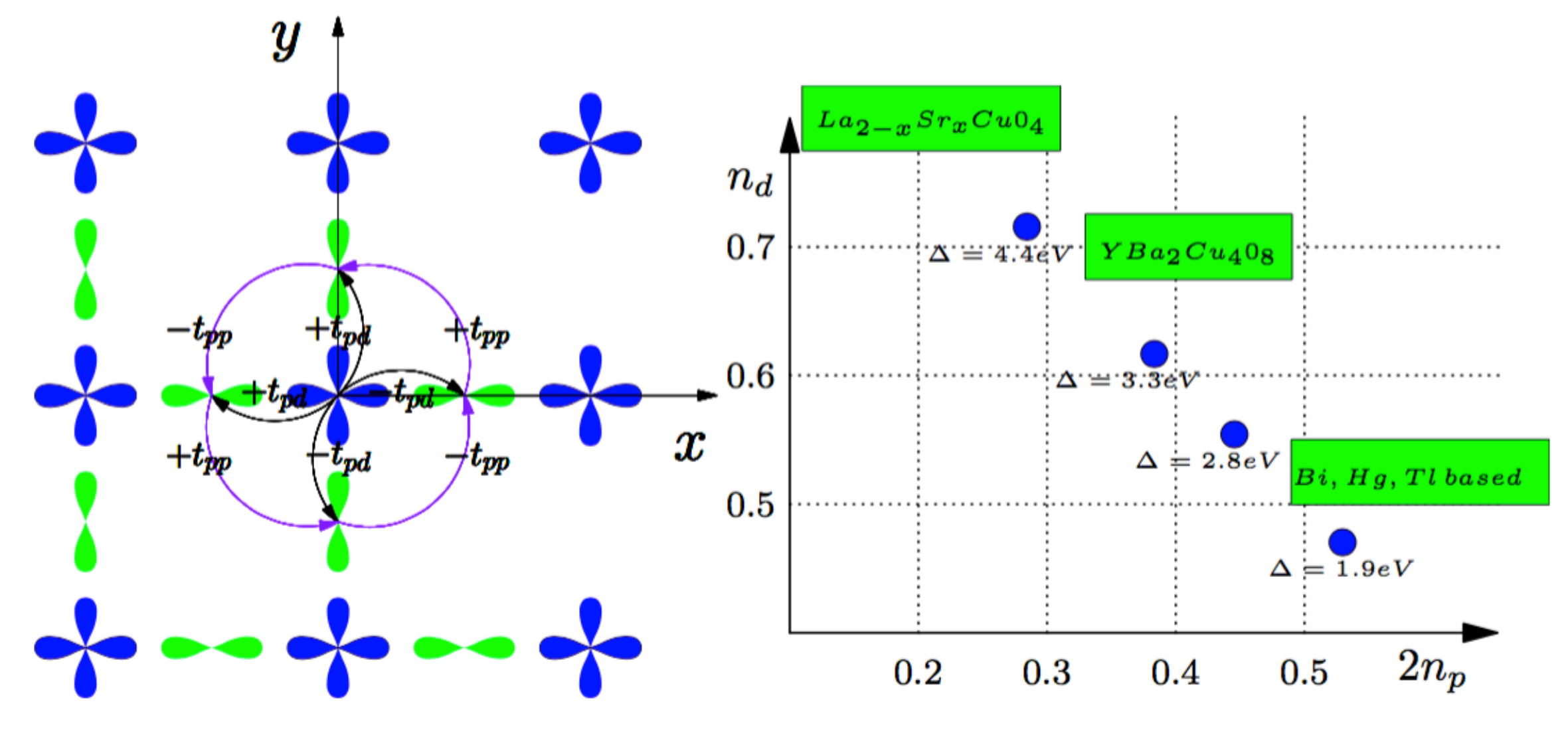}
\caption{ 
 (Color online) (Left) Schematic view of the CuO$_2$ plane %copper-oxygen planes
 of the cuprates. %In blue we represent $
 Cu $3d_{x^2 - y^2}$ orbitals are represented in blue, and
 %while green orbitals are 
 $O$ $2p_x$ and $2p_y$ orbitals in green.
 %We use the reference frame defined by the two axis in the figure.
 The curve connectors represent the hopping, and the labels
 define the sign rule. (Right) Density of holes  around the $d$ and the
 $p$ sites, $n_d$ and $2n_p$ as a function of $\Delta$. 
 Results computed from AFQMC are given by blue circles.
% The blue circles represent results of our simulations. 
The green boxes %{\color{blue} 
are positioned %in such a way 
to
 indicate the typical values of $n_d$ and $n_p$
observed in families of cuprate materials.
\cite{natcommcuprates}}
\label{fig:model}
\end{center}
\end{figure}

This three-band Hubbard model includes the %Copper 
Cu $3d_{x^2 - y^2}$ orbital together with
the %Oxygen 
O $2p_x$ and $2p_y$  orbitals. Most parameter values of the Hamiltonian 
can be derived by \emph{ab initio} methods  for real materials
with reasonable reliability. Among these the charge transfer energy $\Delta$
has been found to vary substantially across different families of cuprate materials, as illustrated in Fig.~\ref{fig:model}. Furthermore, 
%{\color{blue} 
it is known that \emph{ab initio} computations to  determine its value %do pose 
often have difficulties %problems
\cite{PhysRevB.78.035132,PhysRevB.39.9028}. 
This parameter is important because it directly controls the hole density on the Cu 
%and O 
sites,
%and 
which is seen to be 
%experiments indicate is 
anticorrelated with the superconductiong critical temperature
\cite{RUAN20161826,Jurkutat_NMR,Rybicki2016,Weber_chargetransfer}.
%\REMARKS{check above sentence} 
Here we 
investigate the ground-state properties of the parent compound as a function of  $\Delta$,
using state-of-the-art quantum Monte Carlo calculations. 
The calculations are highly accurate, and
benchmark quality results are obtained on the magnetic correlations and charge gaps in the ground state.
We find
that a quantum phase transition occurs at $\Delta \sim 3$eV between a paramagnetic metal and
an antiferromagnetic insulator. 

The Hamiltonian of the three-band Hubbard model is 
\begin{equation}
\label{3bands:ham}
\begin{split}
& \hat{H} = \varepsilon_d \sum_{i, \sigma} \hat{d}^{\dagger}_{i,\sigma} \hat{d}^{}_{i,\sigma} 
+ \varepsilon_p \sum_{j, \sigma} \hat{p}^{\dagger}_{j,\sigma} \hat{p}^{}_{j,\sigma} + \\
& \sum_{<i,j>, \sigma} t_{pd}^{ij} \left(\hat{d}^{\dagger}_{i,\sigma} \hat{p}^{}_{j,\sigma} + h.c \right)
+  \sum_{<j,k>, \sigma} t_{pp}^{jk} \left(\hat{p}^{\dagger}_{j,\sigma} \hat{p}^{}_{k,\sigma} + h.c \right)
\\
& + U_d  \sum_{i} \hat{d}^{\dagger}_{i,\uparrow} \hat{d}^{}_{i,\uparrow} \hat{d}^{\dagger}_{i,\downarrow} \hat{d}^{}_{i,\downarrow} + U_p \sum_{j} \hat{p}^{\dagger}_{j,\uparrow} \hat{p}^{}_{j,\uparrow} \hat{p}^{\dagger}_{j,\downarrow} \hat{p}^{}_{j,\downarrow}\,.
\end{split}
\end{equation}
%In order to introduce the model, we refer to Fig.~\ref{fig:model}, giving a schematic pictorial
%representation of a $Cu O_2$ plane. 
A   %schematic 
pictorial representation of the CuO$_2$ plane is given in  Fig.~\ref{fig:model}. 
We will measure lengths in units of the distance between nearest
neighbors Cu sites.  
In %the operator 
Eq.~\eqref{3bands:ham}, the label $i$ runs over the sites %$\vec{r}_{\rm Cu}$ 
${\mathbf r}_{\rm Cu}$ of a square lattice
$\mathbb{Z}^2$ %made by the positions 
of Cu atoms.
% indicated bythe blue $d_{x^2 - y^2}$ orbitals in Fig.~\ref{fig:model}.
The labels $j$ 
and $k$ run over the positions of the O atoms, shifted with respect
to the Cu sites, %$\vec{r}_{\rm O} = \vec{r}_{\rm Cu} + 0.5 \, \hat{x}$ 
${\mathbf r}_{\rm O} = {\mathbf r}_{\rm Cu} + 0.5\,{\mathbf l}$,
where the unit vector ${\mathbf l}$ is $\hat{x}$
for the $2p_x$ %orbitals 
and $\hat{y}$
%$\vec{r}_{\rm O} = \vec{r}_{\rm Cu} + 0.5 \, \hat{y}$
for the $2p_y$ orbitals.
% as depicted in 
%green  in Fig.~\ref{fig:model}.
%$p$ orbitals in Fig.~\ref{fig:model} depict.
%\REMARKS{Changed vec notation to avoid double 'vec' 'hat' notation in the def of spin corr bbelow - pls
%check to ensure consistency}
The model is formulated in terms of holes: %for example, 
e.g., $\hat{d}^{\dagger}_{i,\sigma}$
creates a hole on the $3d_{x^2 - y^2}$ orbital at site $i$ with spin $\sigma = \uparrow$ or $\downarrow$.
The first two terms %in the hamiltonian contain 
define a charge transfer
energy $\Delta \equiv \varepsilon_p - \varepsilon_d$, representing the energy needed for a hole
to move from a $3d_{x^2 - y^2}$ %orbital 
to %an oxygen 
a $p$ orbital.
The second two terms describe hopping between orbitals; the hopping amplitudes $|t_{pd}^{ij}|=t_{pd}$ and $|t_{pp}^{jk}|=t_{pp}$, with sign convention as illustrated in Fig.~\ref{fig:model}.
% are expressed in terms
%of two parameters, which we will denote $t_{pd}$ and $t_{pp}$,
%and the dependence on the sites is simply a sign factor, which is detailed in  Fig.~\ref{fig:model}.
Finally, the last two terms represent the on-site repulsion energies,
double-occupancy penalties, as in the Hubbard model.

%\begin{figure}[ptb]
%\begin{center}
%\includegraphics[width=6.0cm, angle=270]{pairing_gap.eps}
%\includegraphics[width=6.0cm, angle=0]{np_np.pdf}
%\includegraphics[width=6.0cm, angle=0]{cropped_composite.pdf}
%\caption{ 
 %(Color online) (Upper panel) Density of holes  around the $d$ and the
% $p$ site, $n_d$ and $2n_p$ as a function of $\Delta$. Also non-interacting
%values are shown. (Lower panel) Same results in the $(n_d, 2 n_p)$-plane.
%The blue circles represent results of our simulations. The green boxes, containing names of families of cuprates,
% are positioned in such a way to give an idea of the
 %values observed in the real materials, and are meant to show
 %that the range of values we are considering is realistic for the cuprates.
%More detailed results for real materials are shown in \cite{natcommcuprates}.}
%\label{fig:npnd}
%\end{center}
%\end{figure}
At half-filling, when 
%the number of holes is equal to the number of Cu atoms in the lattice,
there are equal numbers of holes and Cu atoms in the lattice,
the model describes the parent compound, which is known from experiments to be an insulating antiferrmomagnet.
Adding (removing) holes corresponds to hole (electron) doping. %The e
%Experimental evidence shows that, with hole doping,
%the magnetic order rapidly disappears and a superconducting phase becomes the equilibrium state below the %(high) 
%critical temperature. 
%{\color{blue} 
Experimentally, with hole doping, the magnetic order rapidly melts and superconductivity arises 
which competes or cooperates with several forms of spin and charge order.
Naturally, before addressing the topic of superconductivity in the underdoped regime, it is
%extremely 
important to determine the behavior of the model at half-filling.

The Emery model has been studied using several different numerical approaches:
exact diagonalization \cite{PhysRevLett.105.177401,PhysRevB.87.165144}, cluster perturbation theory \cite{PhysRevB.66.075129},
generalized random phase approximation \cite{PhysRevB.92.195140}, %determinantal 
quantum 
Monte Carlo \cite{PhysRevB.93.155166,PhysRevB.57.11980,Huang1161}, density matrix renormalization group \cite{PhysRevB.92.205112}
and dynamical mean field theory or its cluster generalizations \cite{PhysRevLett.114.016402}.
Here we use the CP-AFQMC method \cite{cpprl1995,AFQMC-lecture-notes-2013}, 
which controls the fermion sign problem with a CP approximation that can be systematically improved via a 
self-consistency procedure \cite{PhysRevB.94.235119}. This approach, which has demonstrated 
consistently high accuracy \cite{hubbard_benchmark, Zheng1155}, represents the state-of-the-art many-body 
computational technology for such a system. Our results provide a detailed  
characterization of the ground state properties and reference data on this model at half-filling. 
Furthermore, our calculations establish unambiguously the existence of a metal-insulator transition as a function of 
the charge transfer energy.

%\REMARKS{I commented out two paragraphs. Make sure we don't lose the references  WE ARE OK
%\cite{PhysRevB.93.155166,PhysRevB.39.9028,PhysRevB.78.035132}
%}

Most parameters in the Hamiltonian in  Eq.~\eqref{3bands:ham} have ``canonical''  values
%for cuprates 
obtained from band structure  or other calculations. We will use 
%Our starting point is 
a set of parameters obtained for La$_2$Cu O$_4$, the parent compound of
the lanthanum family of cuprates: $\varepsilon_p = -3.2$, $\varepsilon_d = -7.6$, 
$t_{pd} = 1.2$, $t_{pp} = 0.7$, $U_p = 2$, and $U_d = 8.4$ (all in units of eV).
%This set corresponds to a charge transfer energy $\Delta = 4.4$\,eV. 
The charge transfer energy, however, entails more uncertainty. The set above gives $\Delta = 4.4$\,eV,
but theoretical arguments based on double counting corrections \cite{PhysRevB.78.035132}  would imply a significant reduction to this value.
%{\color{blue}
%A recent %mean field 
%investigation %of the Emery model by ourselves 
%by us \cite{adampaper} has shown that, 
Within generalized Hartree-Fock (GHF),  a %spectacular  
strong dependence of  the ground-state magnetic properties %of the Hartree-Fock ground state 
on $\Delta$ is seen \cite{adampaper}.
%\REMARKS{Discuss this sentence: Standard recipes
%to correct for double counting would imply that we have to reduce this value to $\Delta = 1.5$ eV,
%which, as we will show below, dramatically changes the physical properties of the model, 
%already at half-filling.} 
%However, to determine the value of $\Delta$ theoretically is more subtle. 
Moreover, in real materials,
significant variations have been observed in $\Delta$, % are observed in real materials
%and this parameter 
which can be broadly tuned through chemical substitution and strain \cite{PhysRevB.89.094517}. Recent nuclear magnetic resonance experiments \cite{natcommcuprates} have shown that, as a result,
the hole densities on Cu vary, which in turn affects the critical superconducting transition temperature. 
%
% different families of 
%cuprate materials are believed to show significant variations in $\Delta$, as illustrated in the 
%right panel of  Fig.~\ref{fig:model}. 
%Recent nuclear magnetic resonance experiments \cite{natcommcuprates} have shown that, as a result,
%the hole densities on Cu vary, which in turn affect the critical superconducting transition temperature.
In this study, we scan the value of the charge transfer energy from $\Delta=4.4$ to $1.5$\,eV.

We study systems of $N$ holes in an $M = L \times L$ lattice, 
%$L$ being the number of $Cu$ atoms per direction, 
%that is 
i.e., a supercell of Cu$_{M}$O$_{2M}$.
%and we extrapolate the 
%
%% Large supercells are used in the computation, and the results are extrapolated to the thermodynamic limit,
%% keeping fixed the density at half filling: $ N/L^2 = 1$.}
%We performed 
Calculations are performed on systems 
% lattices 
as large as $L=12$, containing $432$ atoms in the supercell. Special care was taken 
%to provide reliable 
in the extrapolations to the thermodynamic limit. 
%$12 \times 12$, containing $432$ atoms in the supercell
%with $M = 144$ holes. 
Several checks were carried out, with rectangular supercell shapes and with different boundary
conditions % , as well as %the use of 
%separate calculations using 
(periodic and twisted). % boundary conditions.
%of a pinning field (see supplementary material), %have been used to check 
Additionally, calculations with a pinning field to break translational symmetry were also done 
in order to verify the robustness of the long-range order.

To compute properties of the ground state $|\Psi_0\rangle$ of the model, we use the CP-AFQMC 
method, which relies on a projection from an initial or trial wave function:  %formula:
\begin{equation}
|\Psi_0\rangle \propto \lim_{\beta \to +\infty} \exp\left( - \beta (\hat{H} - E_0) \right) \, |\psi_T\rangle\,,
\end{equation}
where $E_0$ is %being 
the ground-state energy which is estimated 
adaptively in the process.
% estimation of the ground state energy. 
The method 
realizes the projection with a stochastic dynamics in the manifold of wave functions of independent particles embedded in random external auxiliary fields.
%{\color{blue} 
The trial wave function $|\psi_T\rangle$ plays an important role in the methodology.
It is used to impose an approximate constraint to the random walk, in order to %which is necessary to
control the  fermion sign problem and
keep the computational complexity at ${\mathcal O}(N^3)$. %polynomial in the system size, precisely
To maximize the accuracy and predictive power 
%robustness 
of the 
approach, we use 
a self-consistent scheme  \cite{Mingpu-sc-PRB} to encode the information from the CP-AFQMC as feedback
in generating a new $|\psi_T\rangle$. %and  improve the predictive power of the constraint. 
We measure the order parameter of a broken-symmetry solution 
of the many-body Hamiltonian with pinning fields. 
A trial wave function is generated using GHF  \cite{adampaper}.
The  CP-AFQMC calculation with 
this  $|\psi_T\rangle$ obtains the density matrix, which is then fed into another GHF 
calculation with renormalized Hamiltonian parameters ($\Delta$ and $U_d$) that are tuned to 
minimize the difference between the density matrix it produces and that from the CP-AFQMC. The new 
GHF solution is then used in a new CP-AFQMC calculation and the process is iterated untill convergence.
This approach has been shown to give very accurate results in a variety of correlated systems
including the one-band Hubbard model \cite{Mingpu-sc-PRB,hubbard_benchmark,Zheng1155}.
%

%{\color{blue}
In Table~\ref{energy::tab} we show the computed ground-state energy per unit cell as a function
of $\Delta$. 
% for $L = 12$, which can provide useful benchmark. 
The results
are obtained as an average over twist angles in the boundary conditions.
The complex phase arising from the twist boundary condition is handled 
straightforwardly \cite{Chang-CP-phase-PhysRevB.78.165101}.
The computed energy is robust with  respect to $|\psi_T\rangle$; the mixed estimate  \cite{CP-long-PhysRevB.55.7464}
is used and no self-consistency iteration is necessary for these results.
The system is large enough such that any residual finite-size effects 
%and dependence on twist 
%parameter are consistently below $10^{-3}$ $eV$, which allowed us to use as
%few as $4$ twist angles for our average.%}
are expected to be comparable to the statistical error bar. 
This was estimated by select calculations with even larger supercell sizes.
%\REMARKS{check these!}
These results should provide valuable benchmark in future studies of the Emery model. 

\begin{table}[ptb]
\caption{Energy per unit cell as a function of the charge-transfer energy.
The values are based on calculations in $12\times 12$ supercells with 
twist-averaging.
%\REMARKS{Check above. Sounds like we think the difference between these and TDL vales are 0.001?
%If ao, we should strengthen the statement}  
}
\label{energy::tab}
\begin{tabular}{llllll}
\hline
\hline
$\Delta \, (eV)$ & 1.9 & 2.8  &  3 & 3.3  & 4.4 \\
\hline
$E/M \, (eV)$ & -10.082(8) & -9.639(2)  &  -9.556(2)  & -9.437(2) & -9.071(6) \\
\hline
\hline
\end{tabular}
\end{table}

\begin{figure}[ptb]
\begin{center}
\includegraphics[width=6.5cm, angle=270]{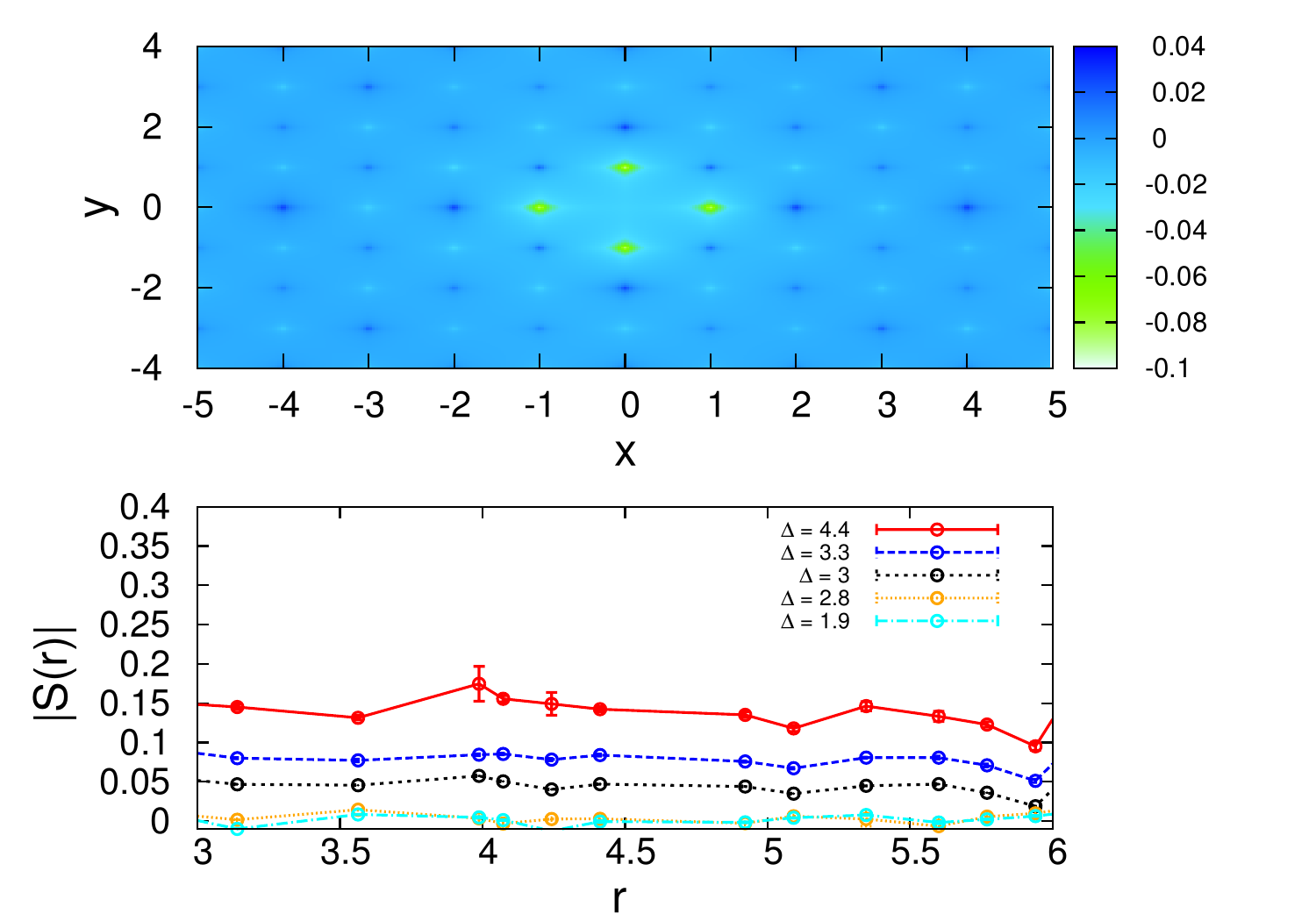}
\caption{ 
(Color online) Computed ground-state magnetic properties.
%spin correlation functions. 
The upper panel shows a color plot of the
result of the spin correlation function at $\Delta = 4.4$ eV 
for a $12\times 12$ supercell. % where the AFM is stable. 
The lower panel shows the
order parameter at asymptotic distances for a sequence of  
% $|S(\vec{r})| = |C_{S}(\vec{r})|^{1/2}$ 
%in the long-range regime for 
%some
 values of the charge transfer energy.} % $\Delta$.}
\label{fig:spin}
\end{center}
\end{figure}

Magnetic properties are presented in Fig.~\ref{fig:spin}.
We measure
%measured by 
spin correlation functions of the form 
%$C_{S}(\vec{r}) = \langle \vec{\hat{S}}(0) \cdot \vec{\hat{S}}(\vec{r})  \rangle$, 
$C_{S}({\mathbf r}) = \langle \hat{\mathbf S}({\mathbf 0}) \cdot \hat{\mathbf S}({\mathbf r})  \rangle$, 
where %, for example if $\vec{r}$ is a Copper site, say $i$,
the spin operator is defined as usual:
$\hat{\mathbf S}({\mathbf r}) = \frac{1}{2} \sum_{\sigma,\sigma^{\prime}}
{\bm \sigma}_{\sigma,\sigma^{\prime}} \, \hat{d}^{\dagger}_{i,\sigma}
\hat{d}^{\dagger}_{i,\sigma^{\prime}}$, with ${\bm \sigma}_{\sigma,\sigma^{\prime}}$ denoting 
elements of the Pauli matrices, %Our results are shown in Fig.~\ref{fig:spin}.
and the expectation $\langle \cdots \rangle$ is with respect to the many-body ground state $|\Psi_0\rangle$,
which requires back-propagation \cite{CP-long-PhysRevB.55.7464}.
%\begin{figure*}
%\end{figure*}
The upper panel is a color plot
of $C_{S}({\mathbf r})$ for $\Delta = 4.4$ eV. 
%It is evident that, while
The correlation function is seen to vanish on the $p$ sites, where no magnetism
is observed. On the other hand,
long-range antiferromagnetic (AFM) order is evident
on the Cu atoms. The lower panel shows the order parameter, 
$|S({\mathbf r})| \equiv |C_{S}({\mathbf r})|^{1/2}$ %in the long-range regime of $|{\mathbf r}|$ 
for  $|{\mathbf r}|\geq 3$ 
as the values of the charge transfer energy $\Delta$ is varied.
%At  $\Delta \geq 3$ eV,
%We see that, while we have 
A non-zero AFM order parameter is seen 
for $\Delta \geq 3$\,eV,
%while 
which becomes 
compatible with
zero for $\Delta \leq 2.8$\,eV,
%below $3$ eV, 
signaling the presence of a phase transition at  $\Delta \sim 3$\,eV. 
%leading to a non-magnetic ground state.
%\begin{figure*}
\begin{figure}[ptb]
\begin{center}
\hspace*{-1cm}
\includegraphics[width=6.5cm, angle=270]{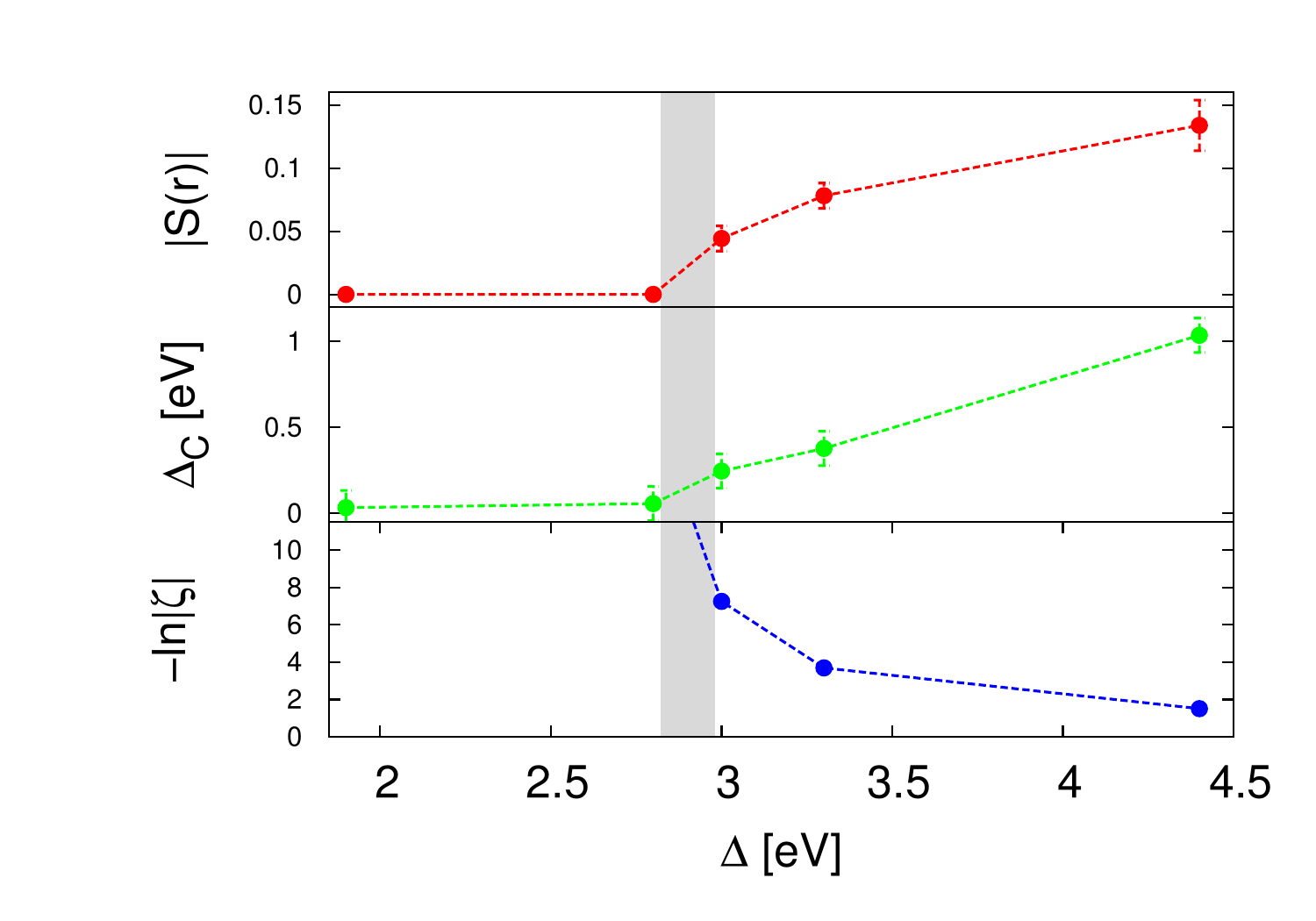}
\caption{%{\bf{Density and spin structure factors}}. 
(Color online) Metal-insulator transition as a function of the charge transfer energy $\Delta$. 
Three different signatures are computed:
%(Upper panel) 
antiferromagnetic order parameter $|S(\vec{r})|$ (upper panel);
%as a function of the charge transfer 
%energy $\Delta$. (Middle panel) 
charge gap $\Delta_C$  
as defined in Eq.~\eqref{eq:chgap} %, as a function of $\Delta$. 
(middle panel);
%(Lower panel) 
logarithm of the %estimator of 
localization measure in Eq.~\eqref{resta} %, as a function of $\Delta$.
(lower panel).
The shaded area indicates %represents 
the phase transition region.
}
\label{fig:phasetransition}
\end{center}
\end{figure}
%\end{figure*}
The asymptotic value of the AFM order parameter (taken as an average over 
%large $|{\mathbf r}|$)  
$|{\mathbf r}|\geq 3$) 
%\REMARKS{changed to above from "large r'. Correct?}
is plotted as a function of $\Delta$
%shown 
in the upper panel of Fig.~\ref{fig:phasetransition}.

%We are naturally induced to study what happens to other
To further examine the 
properties of the system as $\Delta$ becomes smaller, % than $3$\,eV,
we probe the electrical conductivity in the ground state. 
Following Resta and Sorella \cite{PhysRevLett.82.370},
we compute the complex-valued localization measure of the holes: 
\begin{equation}
\label{resta}
\zeta = \left\langle \Psi_0 \, | e^{i \frac{2\pi}{L} \hat{X}} | \, \Psi_0 \right\rangle\,, 
\end{equation}
where, %focussing, for example, on the $x$ direction, 
without loss of generality, we have chosen the quantum mechanical position operator 
${\hat{X}} = {\hat{x}}_1 + \dots + {\hat{x}}_N$ to be along the $x$-direction. %the of the system. 
The quantity $\zeta$, which is related to the quantum metric tensor,
has a 
%We comment that this quantity carries an extremely interesting 
geometrical interpretation and plays an important role in the 
%being related to the quantum metric tensor, and plays a very important role in the 
modern theory of electric polarization. 
A non-zero value of $|\zeta|$ 
for large number of holes implies a localized many-body ground
state and thus an insulator, while a vanishing $|\zeta|$  %vanishes if the system is a conductor.
indicates a delocalized ground state and a conductor.
The dependence of $|\zeta|$ on  $\Delta$  is shown in the 
lower panel of Fig.~\ref{fig:phasetransition}. The result
is consistent with a phase transition from an 
antiferromagnetic, insulating ground state at $\Delta \geq 3$ eV
to a non-magnetic metal at smaller values of the charge-transfer 
energy.
To our knowledge, our calculations here represent one of the first computations  of Eq.~\eqref{resta}
with an advanced many-body method 
in a strongly correlated physical systems whose ground state is unknown. 

We also compute the charge gap of the system
%which is defined in terms of  addition/removal energies:
%ground state energis of systems with $N$, $N+1$ and $N-1$ holes as:
\begin{equation}
\label{eq:chgap}
\Delta_C = E(N+1) + E(N-1) - 2 E(N)\,,
\end{equation}
where $E(N)$ is the ground-state energy at half-filling, while $E(N\pm 1)$ denotes the ground-state energies 
of the system with one hole added/removed. 
%{\color{blue}
The gap is a central quantity %in many-body physics, and it 
which can be directly measured in photoemission spectroscopy experiments.
%The calculation of the gap is prone to subtle size effects. 
Its calculation can be challenging because of finite-size and shell effects arising from the 
non-interacting part of the Hamiltonian. We use 
a scheme  \cite{ettoreGAP} utilizing twist averaging 
%can reduce shell effects arising from the non-interacting hamiltonian and 
to accelerate convergence to the thermodynamic limit.
%In our calculations, w
We find that the dependence on the twist parameter is rather weak here, allowing 
%us to 
%the use of  % $4$ twist
converged results with only a handful of twist
angles in our measurement. A subtlety %role is also played by 
also exists in the choice of trial wave functions for the $(N \pm 1)$
systems. As mentioned before, we build %the trial wave function 
$|\psi_T\rangle$
through a self-consistent procedure providing
a GHF Hamiltonian with renormalized parameters. %choose to 
By using the same mean-field Hamiltonian
to generate the $|\psi_T\rangle$'s %trial wave functions 
for $(N-1)$, $N$ and $(N+1)$-systems, we see better 
error cancellation in tests on smaller systems, and adopt this procedure in the calculation of gaps. 
 % holes. %}
%we perform a twist averaging operation, which is known \cite{ettoreGAP} to dramatically reduce shell effects arising from the non-interacting hamiltonian.
%\REMARKS{say a couple more sentences about what we do? (maybe trial wf? how many twists? etc)}
The result is shown in the middle panel of Fig.~\ref{fig:phasetransition}. %which  shows 
A finite charge gap is seen for $\Delta\geq 3$\,eV, which vanishes at smaller $\Delta$.
%{\color{blue} 
%We observe that, for $\Delta = 4.4$ $eV$, the value %we find 
%is slightly smaller
%than the experimental gap for La$_2$Cu$_4$ \cite{PhysRevB.41.11657,PhysRevB.42.10785,PhysRevB.43.7942,PhysRevLett.69.1109}.
We observe that, for $\Delta = 4.4$ $eV$, the computed gap value is slightly smaller than the experimental gap for La$_2$CuO$_4$  of $1.5-2$ eV \cite{PhysRevB.41.11657,PhysRevB.42.10785,PhysRevB.43.7942,PhysRevLett.69.1109}, but in reasonable agreement given the uncertainties in the choice of Hamiltonian parameters, especially the precise value of $\Delta$. 
% The discrepancy 
%could be due to the actual choice of the set of
%is most likely a consequence of the details of the 
%model parameter values we have adopted.
%\REMARKS{check above statement}
%, or to limitations of the methodology in accurately computing tiny energy differences. %}
%that also the behavior of the charge gap is 

The three independent signatures shown in Fig.~\ref{fig:phasetransition}, the AFM correlation function, the localization measure, and the 
charge gap, all point to a 
consistent picture of the ground state,
with a phase transition from an insulating to a metallic ground state at a charge transfer energy of $\Delta \sim 3$\,eV.
%
%Finally, it is extremely interesting to study non-local charge correlations,

\begin{figure}[ptb]
\begin{center}
\includegraphics[width=8cm, angle=0]{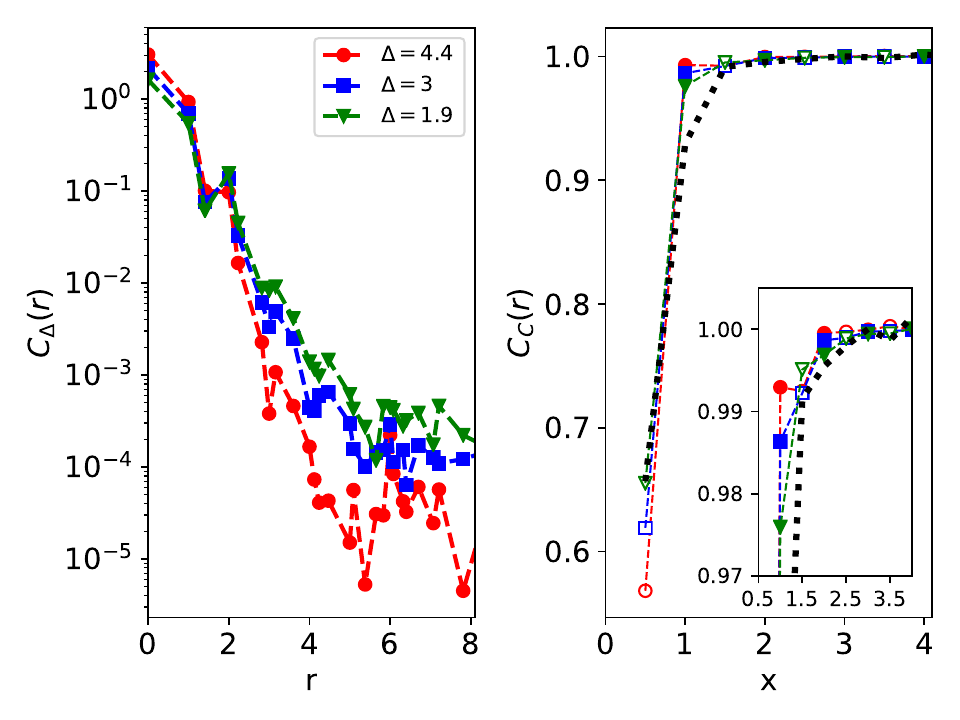}
\caption{ 
(Color online)  
(Left panel) Distance dependence of the $d$-wave pairing correlation 
function $C_{\Delta}({\mathbf r})$  for a few values of the
charge transfer energy.
(Right panel) Density correlation function $C_{C}({\mathbf r})$ 
plotted along the Cu-O bond.
We use open symbols for correlations involving one $d$ orbital and one
$p_x$, while solid symbols indicate $d$-$d$ correlations.
The non-interacting result, which does not depend
on $\Delta$, is also shown %for comparison as a 
(black dotted line) for reference. The inset is a zoom in, for $x > 0.5$. %}
}
\label{fig:density}
\end{center}
\end{figure}

%{\color{blue}
We also investigate the charge density and correlation functions, and the $d$-wave pairing correlations. 
In the right panel of Fig.~\ref{fig:model}, the computed hole densities on the Cu and O sites are shown 
for four different $\Delta$ values spanning the transition. 
%which can be addressed computing the 
Similar to the spin correlation function, we define the charge correlation:
$C_{C}({\mathbf r}) = \langle {\hat{n}}({\mathbf 0}) {\hat{n}}({\mathbf r})  \rangle /  \langle {\hat{n}}({\mathbf 0}) \rangle \langle {\hat{n}}({\mathbf r}) \rangle $, 
where 
the density operator is,  for Cu sites,
$ {\hat{n}}({\mathbf r}) = \sum_{\sigma} \, \hat{d}^{\dagger}_{i,\sigma}
\hat{d}^{\dagger}_{i,\sigma} $, and similarly for the O sites.
The pairing correlation function is defined as:
$C_{\Delta}({\mathbf r}) = \langle {\hat{\Delta}}({\mathbf 0}) {\hat{\Delta}^{\dagger}}({\mathbf r})  \rangle$
where the $d$-wave pairing operator ${\hat{\Delta}}({\mathbf r})$ is defined as in \cite{PhysRevB.57.11980}.
The results are shown in Fig.~\ref{fig:density}. 
%in the right panel we plot
%the pairing correlations as a function of distance, while the lower panel shows the density correlations
%along the Cu-O bond direction.

%\begin{figure*}
%\end{figure*}

% new parag?
%The computed charge correlation is shown In Fig.~\ref{fig:density}.
% we show our results for the charge correlation function $C_{C}(\vec{r})$. The coordinate system is defined in Fig.~\ref{fig:model}, half-integer coordinates indicating Oxygen sites, while
%integer ones denoting Copper sites. 
%The upper panel is a color plot
%of $C_{C}({\mathbf r})$ for $\Delta = 4.4$ eV. 

It is clear that there is no charge and pairing long-range order in the system at half-filling, as expected.
The density correlation function displays only a very short-range repulsive exchange-correlation hole.
As $\Delta$ is increased, the correlation between nearest-neighbor $d$ and $p$ orbitals
decreases while the nearest-neighbor $d$-$d$ correlation increases, another clear
signature of AFM. For the smallest  $\Delta$ , the nearest neighbors $d$-$p$ correlation
is almost identical to the non-interacting result, while the nearest neighbor $d$-$d$ is slightly higher.
%probably due to a 
%which we tendency towards local antiferromagnetism.

Observing the distance dependence of the pairing correlation, %$C_{\Delta}({\mathbf r})$, we notice 
we see that, at very short range, the correlations
increase with $\Delta$, %again 
likely due to the tendency for antiferromagnetic correlations. %On the other hand, 
At longer range, the opposite tendency is seen, %: % becomes evident: 
with the pairing correlations %beyond nearest neighbor distance 
increasing as  $\Delta$ is decreased.
%if we decrease $\Delta$. This is a very interesting result, which 
This result is consistent with the experimental
evidence \cite{RUAN20161826,Jurkutat_NMR,Rybicki2016,Weber_chargetransfer}
%\REMARKS{refs?}
that the charge-transfer energy is anticorrelated with the superconducting critical temperature.
%The result suggests a 
It suggests a picture of local tendency towards AFM order which  
allows the system to build $d$-wave pairs that become more correlated 
once the holes become more 
delocalized. %, allows the system to build d-wave pairs.
%It will be crucial to investigate the fate of those correlations when doped systems will be studied.
Clearly it will be very interesting and important in the future to investigate the behavior of these correlations with doping.

In summary, we performed an extensive study of the ground state of the Emery model at half-filling
using a cutting-edge many-body technique, CP-AFQMC.
%quantum Monte Carlo methods. 
The favorable computational scaling of the algorithm allowed
us to study supercells as large as $12 \times 12$ which, together with twist averaging, makes it
possible to access properties at the thermodynamic limit. We investigated the role
of the charge transfer energy $\Delta$, whose value is less well determined and 
appears to vary across different families of 
cuprate materials.
%cannot be predicted ab-initio due to
%a double counting issue. 
Accurate results on the spin correlation functions, the localization or conductivity measure, % \eqref{resta},
and the charge gap 
%\eqref{eq:chgap} 
are computed versus $\Delta$ for a set of canonical Hamiltonian parameters.
Ground-state energies, 
charge densities and correlation functions, and pairing correlations are also determined.
%We also
%studied charge correlation functions which displayed very interesting non local
%correlations among holes, with a very intriguing interplay with spin correlations.
%The results show %data are consistent with 
%{\color{blue} Also, a tendency of the d-wave pairing to be anticorrelated with $\Delta$ emerges.}
%
 The tendency of $d$-wave pairing is seen to increase as  $\Delta$ decreases.
% to be anticorrelated with $\Delta$ emerges.
%\REMARKS{check above sentence}
Our results establish unambiguously 
a phase transition in the ground state of this fundamental model connecting
an antiferromagnetic insulator to a non-magnetic metal as $\Delta$ is decreased to $\sim 3$\,eV.
%which becomes stable when $\Delta$ is
%smaller that $3$ eV. 

%{\color{blue} 
We thank the Simons Foundation and NSF (Grant No. DMR-1409510) for their support.
Computing was carried out at the Extreme Science and Engineering Discovery Environment (XSEDE),
which is supported by National Science Foundation grant number ACI-1053575, and the High Performance
Computational facilities
facilities at William and Mary.
The Flatiron Institute is a division of the Simons Foundation.
%}

%merlin.mbs apsrev4-1.bst 2010-07-25 4.21a (PWD, AO, DPC) hacked
%Control: key (0)
%Control: author (8) initials jnrlst
%Control: editor formatted (1) identically to author
%Control: production of article title (-1) disabled
%Control: page (0) single
%Control: year (1) truncated
%Control: production of eprint (0) enabled
%

%\bibliography{hub.bib}

\end{document}